\documentclass[aps,pra,onecolumn,superscriptaddress,showpacs,showkeys]{revtex4}
\usepackage[]{graphicx}
\begin{document}

\title{Brownian motion in a non-homogeneous force field and photonic force microscope}
\author{Giorgio Volpe}
\affiliation{ICFO - Institut de Ciencies Fotoniques,
Mediterranean Technology Park, 08860, Castelldefels (Barcelona),
Spain}
\author{Giovanni Volpe}
\affiliation{ICFO - Institut de Ciencies Fotoniques,
Mediterranean Technology Park, 08860, Castelldefels (Barcelona),
Spain}
\author{Dmitri Petrov}
\affiliation{ICFO - Institut de
Ciencies Fotoniques, Mediterranean Technology Park, 08860,
Castelldefels (Barcelona), Spain} \affiliation{ICREA -
Instituci\'{o} Catalana de Recerca i Estudis Avan\c{c}ats, 08010,
Barcelona, Spain}
\date{\today}

\begin{abstract}
The Photonic Force Microscope (PFM) is an opto-mechanical
technique based on an optical trap that can be assumed to probe forces in microscopic systems.
This technique has been used to measure forces in the range of pico- and femto-Newton, assessing the mechanical
properties of biomolecules as well as of other microscopic systems.
For a correct use of the PFM, the force field to measure has to be
invariable (homogeneous) on the scale of the Brownian motion of
the trapped probe.
This condition implicates that the force field must be conservative, excluding
the possibility of a rotational component.
However, there are
cases where these assumptions are not fulfilled
Here, we show how to improve the PFM technique in order to be
able to deal with these cases.
We introduce the theory of this enhanced PFM and we propose a concrete analysis workflow
to reconstruct the force field from the experimental time-series of the probe position.
Furthermore, we experimentally verify some particularly important cases, namely the case of a
conservative or rotational force-field.
\end{abstract}

\pacs{05.40.Jc, 87.80.Cc, 07.10.Pz, 47.61.-k}
\keywords{Brownian motion, Photonic Force Microscope, Singular points, Stability analysis}
\maketitle

\section{Introduction}\label{sec:intro}

A focused optical beam - an optical tweezers - permits one to manipulate a wide range of
particles - including atoms, molecules, DNA fragments, living
biological cells, and organelles within them - with applications
to many areas - such as molecular biophysics, genetic
manipulation, micro-assembly, and micro-machines
\cite{Ashkin1970,Ashkin1986,Neuman2004}. One of the most exciting applications has been
the possibility to investigate and engineer the mechanical
properties of microscopic systems - using, for example, optical
traps as force transducers for mechanical measurements in
biological systems
\cite{Ashkin1990,Block1990,Svoboda1993,Kuo1993,Finer1994}.

In the early 90s various kinds of scanning probe microscopy were
already established.
The Scanning Tunneling Microscope
(STM) \cite{Binnig1982} permits one to resolve at the atomic level
crystallographic structures \cite{Binnig1983} and organic
molecules \cite{Smith1990}. The Atomic Force Microscope (AFM) \cite{Binnig1986}
has been successfully employed to study biological and nano-fabricated
structures, overcoming the diffraction limit of
optical microscopes. Furthermore, they developed from pure imaging
tools into more general manipulation and measuring tools on the
level of single atoms or molecules. However, all these techniques
required a macroscopic mechanical device to guide the probe.

A new kind of scanning
force microscope using an optically trapped dielectric microsphere as a probe was proposed in
\cite{Ghislain1993,Ghislain1994}. This technique was later called
Photonic Force Microscope (PFM) \cite{Florin1997}.  In a typical
setup, the probe is held
in an optical trap, where it performs random movements due to its
thermal energy. The analysis of this thermal motion provides
information about the local forces acting on the probe.
The three-dimensional probe position can be recorded
through different techniques which detect its forward or backward
scattered light. The most commonly used are a
quadrant photodiode, a position sensing detector, or a
high-speed video-camera \cite{Neuman2004}. The PFM provides the capability of
measuring forces in the range from femto- to pico-Newton. These
values are well below those achieved with techniques
based on micro-fabricated mechanical cantilevers, such as AFM
\cite{Weisenhorn1989}.

For small displacements of the probe from the center of an optical
trap, the restoring force is proportional to the displacement.
Hence, an optical trap acts on the probe like a Hookeian spring
with a fixed stiffness, which can be characterized with various
methods \cite{Visscher1996,Neuman2004}. The \textit{correlation or
power spectrum method}, in particular, is considered the most
reliable \cite{Visscher1996}, allowing one to determine the trap
stiffness by applying Boltzmann statistics to the position
fluctuations of the probe, relying only on the knowledge of the
temperature and the viscosity of the surrounding medium
\cite{Ghislain1993,Ghislain1994,Florin1997,Rohrbach2002a,Berg-sorensen2004}.

Assuming a very low Reynolds number regime
\cite{Purcell1977,HappelBrenner}, the Brownian motion of the
probe in the optical trap is described by a set of Langevin equations:
\begin{equation}
\label{eqn:ClassicalPFM} \gamma \mathbf{\dot{r}}(t) + \mathbf{K}
\mathbf{r}(t) = \sqrt{2D} \gamma \mathbf{h}(t),
\end{equation}
where $\mathbf{r}(t) = [x(t),y(t),z(t)]$ is the probe position,
$\gamma = 6\pi R \eta$ is its friction coefficient, $R$ is the
probe radius, $\eta$ is the medium viscosity, $\mathbf{K}$ is the
restoring force matrix, $\sqrt{2 D} \gamma \mathbf{h}(t) = \sqrt{2 D} \gamma [h_x(t),h_y(t),h_z(t)]$ is a
vector of independent white Gaussian random processes describing
the Brownian forces, $D = k_B T / \gamma$ is the diffusion coefficient,
$T$ is the absolute temperature, and $k_{B}$ is
the Boltzmann constant. The orientation of the coordinate system
can be chosen in such a way that the restoring force is
independent in the three directions, i.e. $\mathbf{K} =
\mathrm{diag}(k_x,k_y,k_x)$.  In this reference frame
the stochastic differential equations (\ref{eqn:ClassicalPFM})
are separated and without loss of generality
the treatment can be restricted to the $x$-projection of the system.

The autocorrelation function (ACF) of the solution to equations (\ref{eqn:ClassicalPFM})  in each direction reads
\begin{equation}
\label{eqn:ClassicalACF}
<x(t)x(t+\Delta t)> = D \frac{\gamma}{k_x} e^{-k_x |\Delta t|/\gamma},
\end{equation}
where $k_x$ is the trap stiffness. Experimentally the trap
stiffness is found by fitting the theoretical ACF
(\ref{eqn:ClassicalACF}) to the one obtained from the
measurements. Using the Wiener-Khintchine theorem, the power
spectral density (PSD) can now be calculated as the Fourier
transform of the ACF:
\begin{equation}
\label{eqn:ClassicalPSD} P_x(f)
=\frac{D}{2\pi^2 \left(f^2+f_c^2\right)},
\end{equation}
where $f_{c}=k_x /2\pi \gamma $ is the corner frequency.

A constant and homogeneous external force $f_{ext,x}$ acting on the probe produces a shift in its
equilibrium position in the trap. The value of the force can be obtained as:
\begin{equation}
\label{eqn:ClassicalFx}
f_{ext,x} = k_x <x(t)>,
\end{equation}
where $<x(t)>$ is the probe mean displacement from the previous
equilibrium position.

The PFM has been applied to measure forces
in the range of femto- to pico-Newton in many different fields
with exciting applications, for example, in biophysics,
thermodynamics of small systems, and colloidal physics
\cite{Ashkin1990,Block1990,Svoboda1993,Kuo1993,Finer1994,Pralle1998,Menta1999,Pralle2000,
Smith2001,Lang2002,Smith2003,Rohrbach2005a,Volpe2006c,Volpe2006a}

\begin{figure}[tbp]
\includegraphics[width = 8.5cm]{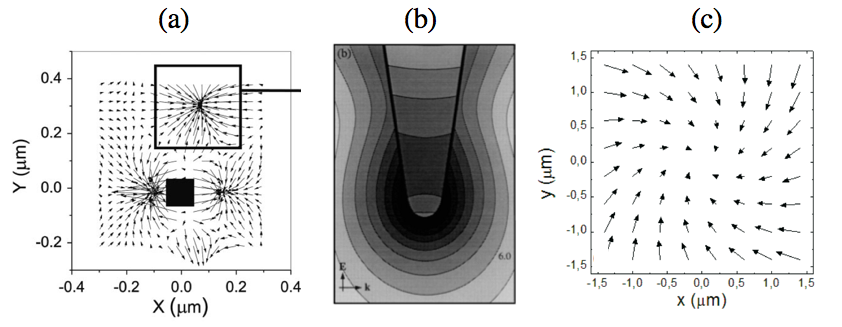}
\caption{ (Color online) Examples of physical systems that produce forces fields that can
not be correctly probed with a classical PFM because they vary on the scale of the Brownian motion of
the trapped probe, i.e. in a range of $10$s to $100$s of
nanometers : (a) forces produced by a surface plasmon polariton in the presence of
a patterned surface on a $50\, nm$ radius dielectric particle (reproduced from \cite{Quidant2005});
(b)  electromagnetic near-field of a $10 \, nm$ wide gold tip in water illuminated
by a $810 \, nm$ monochromatic light beam (reproduced from \cite{Novotny1997}); and (c)
vector force field acting on a $500 \, nm$ radius dielectric particle in the focal plane of a highly focused
Laguerre-Gaussian beam (reproduced from \cite{Volpe2006b}).}
\label{f:examples}
\end{figure}

For a correct use of the PFM, the force field to measure has to be
invariable (homogeneous) on the scale of the Brownian motion of
the trapped probe, i.e. in a range of $10$s to $100$s of
nanometers depending on the trapping stiffness. This condition
implicates that the force field must be conservative, excluding
the possibility of a rotational component. However, there are
cases where these assumptions are not fulfilled as it is
illustrated in Fig. \ref{f:examples}. The field can vary in the
nanometer scale, for example, in the presence of a radiation force
field produced by a surface plasmon polariton \cite{Volpe2006a}.
It can also be non-conservative in the presence of a rotational
force (torque). These effects appear, for example, considering the
radiation forces exerted on a dielectric particle by a patterned
optical near-field landscape at an interface decorated with
resonant gold nanostructures \cite{Quidant2005} (Fig.
\ref{f:examples}(a)); the nanoscale trapping that can be achieved
near a laser-illuminated tip \cite{Novotny1997} (Fig.
\ref{f:examples}(b)); the optical forces produced by a beam which
carries orbital angular momentum \cite{Volpe2006b} (Fig.
\ref{f:examples}(c)); or in the presence of fluid flows
\cite{Giorgio2007}.

In this article, we show how to improve the PFM technique in order to be
able to deal with these cases.
We introduce the theory of this enhanced PFM (section \ref{sec:theory}).
Based on this theoretical understanding, in section \ref{sec:experiments} we propose a concrete analysis workflow
to reconstruct the force field from the experimental time-series of the probe position.
Finally, in section \ref{sec:cases} we present experimental results for some particularly important cases, namely the case of a
conservative or rotational force-field.

\section{Theory}\label{sec:theory}

In the presence of an external force field
$\mathbf{f_{ext}}(\mathbf{r}(t))$, equation
(\ref{eqn:ClassicalPFM}) can be written in the form:
\begin{equation}
\label{e:ForceFieldBM} \gamma \mathbf{\dot{r}}(t) =
\mathbf{f}(\mathbf{r}(t)) + \sqrt{2 D} \gamma \mathbf{h}(t),
\end{equation}
where the total force acting on the probe
$\mathbf{f}\left(\mathbf{r}(t)\right) =
\mathbf{f_{ext}}(\mathbf{r}(t)) - \mathbf{K}\mathbf{r}(t) =
\left[f_x(\mathbf{r}(t)),f_y(\mathbf{r}(t))\right]$ depends on
the position of the probe itself, but does not vary over time. We
reduce our analysis to a bidimensional system, because it is the
most interesting from the applied point of view. However, our
approach can be generalized to the tridimensional case.

The force $\mathbf{f}(\mathbf{r}(t))$ can be expanded in Taylor
series up to the first order around an arbitrary point
$\mathbf{\tilde{r}}$:
\begin{equation}
\label{e:Taylor}
\mathbf{f}(\mathbf{r}(t)) =
\underbrace{
\left[\begin{array}{c}
f_{x}(\mathbf{\tilde{r}}) \\
f_{y}(\mathbf{\tilde{r}}) \\
\end{array}\right]
}_{\mathbf{f_{\tilde{r}}}}
+
\underbrace{
\left[\begin{array}{cc}
\partial_x f_x(\mathbf{\tilde{r}}) & \partial_y f_x(\mathbf{\tilde{r}}) \\
\partial_x f_y(\mathbf{\tilde{r}}) & \partial_y f_y(\mathbf{\tilde{r}}) \\
\end{array}\right]
}_{\mathbf{J_{\tilde{r}}}}
\left( \mathbf{r}(t) - \mathbf{\tilde{r}} \right) + o\left( || \mathbf{r} - \mathbf{\tilde{r}}|| \right),
\end{equation}
where  $\mathbf{f_{\tilde{r}}}$ and $\mathbf{J_{\tilde{r}}}$ are
the zeroth-order expansion, i.e. the force field value at the
point $\mathbf{\tilde{r}}$, and the Jacobian of the force field
calculated in $\mathbf{\tilde{r}}$, respectively.

In a PFM the probe particle is optically trapped and, therefore,
it diffuses due to Brownian motion in the total force field (the
sum of the optical one and the one under investigation). If
$\mathbf{f_{\tilde{r}}} \neq \mathbf{0}$, the probe experiences a
shift in the direction of the force. After a
time has elapsed, therefore, the particle settles down in a
new equilibrium position of the total force field, such that
$\mathbf{f_{\tilde{r}}} = \mathbf{0}$.
As seen in the introduction, the measurement of this shift allows one to evaluate the homogeneous force
acting on the probe in the standard PFM. Assuming,  without loss of
generality, $\mathbf{\tilde{r}}=\mathbf{0}$, the statistics of the
Brownian motion in the surroundings of the equilibrium point can
be analyzed in order to reconstruct
the force field up to its first-order approximation.

\subsection{Brownian motion near an equilibrium position}

The first order approximation to equation (\ref{e:ForceFieldBM})
near a stable force field equilibrium point, $\mathbf{\tilde{r}} =
\mathbf{0}$, is:
\begin{equation}
\label{e:MotionNearEquilibrium} \mathbf{\dot{r}}(t) = \gamma^{-1} \mathbf{J_0} \mathbf{r}(t) +
\sqrt{2 D} \mathbf{h}(t),
\end{equation}
where $\mathbf{r}(t) = \left[x(t), y(t) \right]$,
$\mathbf{h}(t) = \left[h_x(t), h_y(t)\right]$, and $\mathbf{J_0}$
is the Jacobian calculated in the equilibrium point.
According to the Helmholtz theorem,
under reasonable regularity conditions any force field can be
separated into its conservative (i.e. irrotational) and
non-conservative (i.e. rotational or solenoidal) components.
The two components  can be immediately identified
if the coordinate system is chosen such that
$\partial_y f_x(\mathbf{0}) = - \partial_x f_y(\mathbf{0})$. In
this case, the Jacobian $\mathbf{J_0}$ normalized by the friction
coefficient $\gamma$ reads:
\begin{equation}
\label{e:J0}
\gamma^{-1}\mathbf{J_0} =
\left[\begin{array}{cc}
-\phi_x & \Omega \\
-\Omega & -\phi_y \\
\end{array}\right],
\end{equation}
where $\phi_x = k_x/\gamma$ and $\phi_y = k_y/\gamma$, $k_x
= -\partial_x f_x(\mathbf{\tilde{r}})$ and $k_y = -\partial_y
f_y(\mathbf{\tilde{r}})$, and $\Omega = \gamma^{-1} \partial_y
f_x(\mathbf{\tilde{r}}) = -\gamma^{-1} \partial_x
f_y(\mathbf{\tilde{r}})$.
In (\ref{e:J0}) the rotational component, which is invariant under
a coordinate rotation, is represented by the non-diagonal terms of
the matrix: $\Omega$ is the value of the constant angular velocity
of the probe rotation around the $z$ axis due to the
presence of the rotational force field \cite{Volpe2006b}. The
conservative component, instead, is represented by the diagonal
terms of the Jacobian and is centrally symmetric with respect to the
origin. Without loss of generality, we impose that $k_x>k_y$, i.e.
$\phi_x>\phi_y$. This means that the stiffness of the trapping potential is
higher along the $x$-axis.

\begin{figure}[tbp]
\includegraphics[width = 8.5cm]{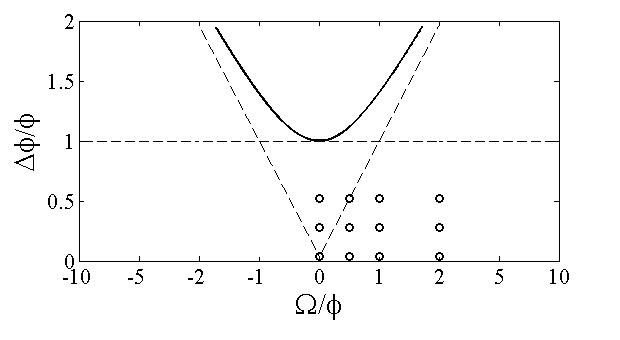}
\caption{(Color online) Stability diagram. Assuming $\phi > 0$,
the stability of the system is shown as a function of the parameters  $\Omega/\phi$ and $\Delta\phi/\phi$.
The regions that satisfy the conditions described in equation (\ref{e:Stability}). The continuos line delimits the stability region.
The dashed lines
represent the $\Delta\phi = |\Omega|$ and $\Delta\phi = \phi$. The dots represent the parameters that are further investigated in Fig.
(\ref{f:force-fields}), (\ref{f:correlations}) and (\ref{f:invariant}).}
\label{f:stability}
\end{figure}

The equilibrium point is stable if:
\begin{equation} \label{e:Stability} \left\{
\begin{array}{*{20}l}
\mathrm{Det}\left(\mathbf{J_0}\right) =  \phi^2 - \Delta\phi^2 + \Omega^2  \ge 0 \\
\mathrm{Tr}\left(\mathbf{J_0}\right) = -2 \phi \le 0 \\
\end{array}
\right. ,
\end{equation}
where $\phi = (\phi_x + \phi_y)/2$ and $\Delta\phi =
(\phi_x - \phi_y)/2$.
The fundamental condition required to achieve stability is
$\phi > 0$. Assuming that this condition is satisfied,
the behavior of the optically trapped probe
can be explored as a function of the parameters $\Omega/\phi$ and $\Delta\phi/\phi$.
The stability diagram is shown in Fig. \ref{f:stability}.
The standard PFM corresponds to $\Delta\phi = 0$
and $\Omega = 0$. When a rotational term is added, i.e. $\Omega \neq 0$ and $\Delta\phi = 0$,
the system remains stable \cite{Volpe2006b}.
When there are no
rotational contributions to the force field ($\Omega = 0$)
the equilibrium point becomes unstable as soon as $\Delta\phi \ge
\phi$. This implicates that $\phi_y < 0$, and therefore
the probe is not confined in the $y$-direction any more.
In the presence of a rotational component ($\Omega \neq 0$) the
stability region becomes larger; the equilibrium point now becomes unstable only for
$\Delta\phi \ge \sqrt{\phi^2-\Omega^2}$.

\begin{figure}[tbp]
\includegraphics[width = 8.5cm]{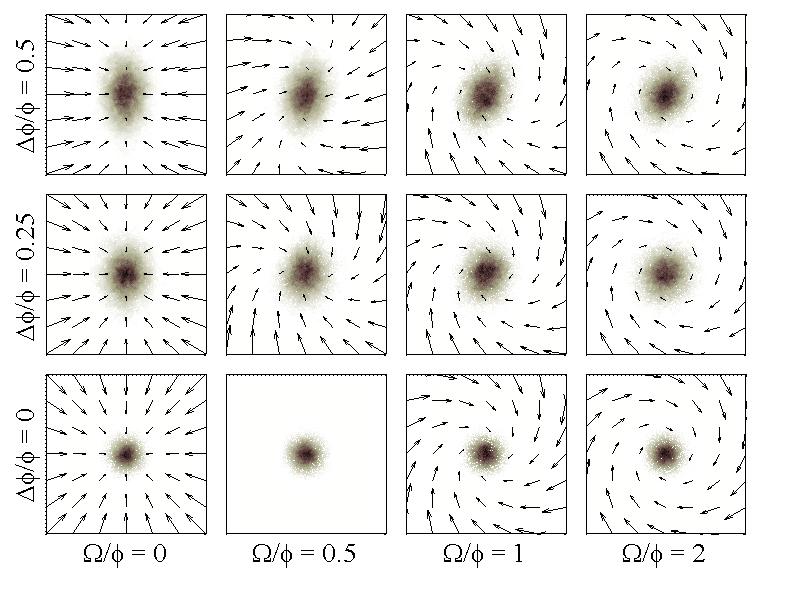}
\caption{(Color online) The arrows show the force field vectors for various values of the parameters $\Delta\phi/\phi$ and
$\Omega/\phi$. The shadowed areas show the probability distribution function (PDF) of the probe position in the corresponding force field.}
\label{f:force-fields}
\end{figure}

Some examples of possible force fields are presented in Fig.
\ref{f:force-fields}.

When $\Omega = 0$ the probe movement can be
separated along two orthogonal directions. As the value of
$\Delta\phi$ increases the probability density function (PDF) of
the probe position becomes more and more
elliptical, until for $\Delta\phi \ge \phi$ the particle is confined only along the
$x$-direction and the confinement along the $y$-direction is
lost.

If $\Delta\phi = 0$, the increase in $\Omega$ induces a
bending of the force field lines and the probe movement along the $x$- and $y$-directions
are not independent any more. For
values of $\Omega \ge \phi$, the rotational component of the force field
becomes dominant over the conservative one. This is
particularly clear when $\Delta\phi \neq 0$: the
presence of a rotational component covers the asymmetry in the
conservative one, since the probability density distribution
assumes a more rotationally-symmetric shape.

\subsection{Enhanced Photonic Force Microscope}

As we already mentioned in the introduction,
the most powerful analysis method
is based on the study of
the correlation functions - or, equivalently, the power spectral density -
of the probe position time-series.
In this subsection, we
derive the correlation matrix first in the
coordinate system considered in the previous subsection,
where the conservative and rotational components are separated.
We, then, derive the same matrix
in a generic coordinate system and identify some
functions that are independent on the choice of the coordinate system.
For completeness, we will also present the
power spectral density matrix.

\subsubsection{Correlation Matrix}

The correlation matrix of the probe motion near an equilibrium
position can be calculated from the solutions of
(\ref{e:MotionNearEquilibrium}), whose eigenvalues are $\lambda_{1,2} =
-\phi \pm \sqrt{\Delta\phi^2 - \Omega^2}$
and whose eigenvectors are $\mathbf{v_{1,2}} = \left[\Omega, \Delta\phi
\pm \sqrt{\Delta\phi^2 - \Omega^2}\right]^T$.

Treating $\mathbf{h}(t)$ as a driving function, the solution of
(\ref{e:MotionNearEquilibrium}) is given by:
\begin{equation}
\label{} \mathbf{r}(t) = \sqrt{2 D} \int_{-\infty}^{t}
\mathbf{W}(t) \mathbf{W}^{-1}(s) \mathbf{h}(s) ds,
\end{equation}
where
\begin{equation}
\label{} \mathbf{W}(t) = \left[\begin{array}{cc}
\Omega & \Omega \\
\Delta\phi + \sqrt{\Delta\phi^2 - \Omega^2} & \Delta\phi - \sqrt{\Delta\phi^2 - \Omega^2} \\
\end{array}\right]
\left[\begin{array}{cc}
e^{\lambda_{1} t} & 0 \\
0 & e^{\lambda_{2} t} \\
\end{array}\right]
\end{equation}
is the Wronskian of the system.

Since we are assuming
$\mathbf{r}(t)$ to be a stationary stochastic process,
the correlation matrix $\left<\mathbf{r}(t+ \Delta
t)\mathbf{r^*}\left(t \right)\right>$ can be obtained by taking the
ensemble average $\left<\mathbf{r}(\Delta
t)\mathbf{r^*}\left( 0 \right)\right>$:
\begin{equation}
\label{} \left<\mathbf{r}(\Delta t)\mathbf{r^h}\left(
0\right)\right> = \left< 2 D \int_{-\infty}^{\Delta t}
\mathbf{W}(\Delta t) \mathbf{W}^{-1}(t') \mathbf{h}(t') dt'
\int_{-\infty}^{0} \mathbf{h^h}(t'')
\mathbf{W}^{-1\mathbf{h}}(t'') \mathbf{W^h}(0) dt'' \right>,
\end{equation}
where the superscript $h$ indicates the hermitian.
Solving this system, we have
\begin{eqnarray}
r_{xx} (\Delta t)
& = &
D \frac{e^{-\phi \left|\Delta t\right|}}{\phi}
\left[
\left(
\frac{
\Omega^2
-
\alpha^2 \Delta\phi^2
}
{\Omega^2-\Delta\phi^2}
-\alpha^2 \frac{\Delta\phi}{\phi}
\right)
\mathcal{C}(\Delta t)
-
\alpha^2 \frac{\Delta\phi}{\phi}
\left(1 - \frac{\Delta\phi}{\phi} \right)
\mathcal{S} (|\Delta t|)
\right] ,
\label{e:rxx}
\\
r_{yy} (\Delta t)
& = &
D \frac{e^{-\phi \left|\Delta t\right|}}{\phi}
\left[
\left(
\frac{
\Omega^2
-
\alpha^2 \Delta\phi^2
}
{\Omega^2-\Delta\phi^2}
+\alpha^2 \frac{\Delta\phi}{\phi}
\right)
\mathcal{C}(\Delta t)
+
\alpha^2 \frac{\Delta\phi}{\phi}
\left(1 + \frac{\Delta\phi}{\phi} \right)
\mathcal{S} (|\Delta t|)
\right] ,
\label{e:ryy}
\\
r_{xy} (\Delta t)
& = &
D \frac{e^{-\phi \left|\Delta t\right|}}{\phi}
\frac{\Omega}{\phi}
\left[
+\mathcal{S} (\Delta t)
+
\alpha^2 \frac{\Delta\phi}{\phi}
\left( \mathcal{C} (\Delta t) + \mathcal{S} (|\Delta t|) \right)
\right] ,
\label{e:rxy}
\\
r_{yx} (\Delta t)
& = &
D \frac{e^{-\phi \left|\Delta t\right|}}{\phi}
\frac{\Omega}{\phi}
\left[
-\mathcal{S} (\Delta t)
+
\alpha^2 \frac{\Delta\phi}{\phi}
\left( \mathcal{C} (\Delta t) + \mathcal{S} (|\Delta t|) \right)
\right] ,
\label{e:ryx}
\end{eqnarray}
where
\begin{equation}
\alpha^2 =  \frac{\phi^2}{\phi^2 + \left(\Omega^2 - \Delta\phi^2 \right)}
\end{equation}
is a dimensionless positive parameter,
\begin{equation} \label{}
\mathcal{C}(t) = \left\{
\begin{array}{cc}
\cos \left( \sqrt{|\Delta\phi^2 - \Omega^2|} t \right) & \Omega^2 > \Delta\phi^2 \\
1 & \Omega^2 = \Delta\phi^2 \\
\mathrm{cosh} \left( \sqrt{|\Delta\phi^2 - \Omega^2|} t \right) & \Omega^2 < \Delta\phi^2 \\
\end{array}
\right.
\end{equation}
and
\begin{equation} \label{}
\mathcal{S}(t) = \left\{
\begin{array}{cc}
\phi \frac{\sin \left( \sqrt{|\Delta\phi^2 - \Omega^2|} t \right)}{\sqrt{|\Delta\phi^2 - \Omega^2|}}  & \Omega^2 > \Delta\phi^2 \\
\phi t & \Omega^2 = \Delta\phi^2 \\
\phi \frac{\mathrm{sinh} \left( \sqrt{|\Delta\phi^2 - \Omega^2|} t \right)}{\sqrt{|\Delta\phi^2 - \Omega^2|}} & \Omega^2 < \Delta\phi^2 \\
\end{array}
\right. .
\end{equation}

\begin{figure}[tbp]
\includegraphics[width = 8.5cm]{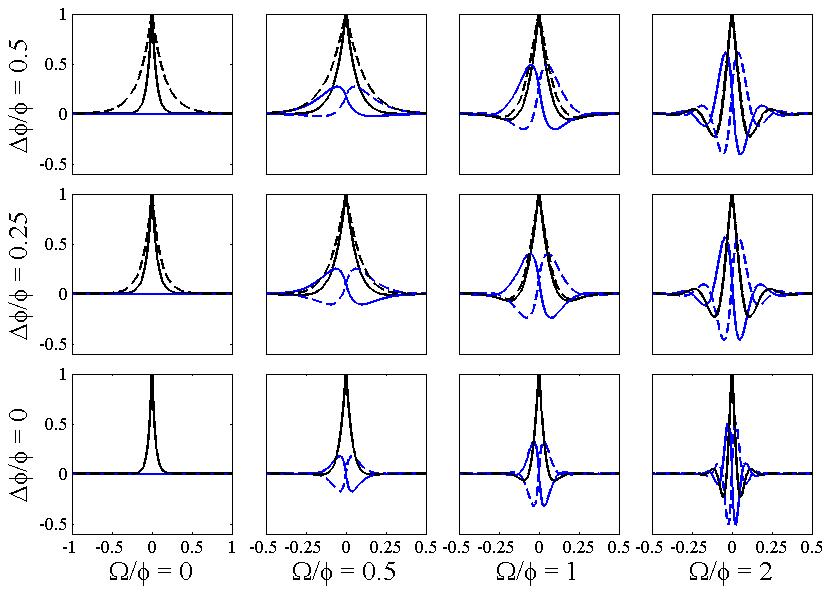}
\caption{(Color online) Auto- and cross-correlation functions for various values of the parameters
$\Delta\phi/\phi$ and $\psi/\phi$: $r_{xx}$ (black continuous line), $r_{yy}$ (black dotted line),
$r_{xy}$ (blue continuous line), and $r_{yx}$ (blue dotted line).}
\label{f:correlations}
\end{figure}

In Fig. \ref{f:correlations} these functions are plotted for different ratios of the
force field conservative and rotational components. Some
cases have already been studied experimentally. For the case
$\Delta\phi = 0$ \cite{Volpe2006b}, the ACFs and cross-correlation functions (CCFs)
are $r_{xx} (\Delta t) = r_{yy} (\Delta t) = D e^{-\phi |\Delta
t|} \cos(\Omega \Delta t)/ \phi$ and  $r_{xy} (\Delta t) = -
r_{yx} (\Delta t)= D e^{-\phi |\Delta t|} \sin(\Omega \Delta
t)/\phi$, respectively. As the rotational component becomes
greater than the conservative one ($\Omega > \phi$), a first
zero appears in the ACFs and CCFs and, as $\Omega$ increases even further, the number of oscillation
grows. Eventually, for $\Omega \gg \phi$
the sinusoidal component becomes dominant. The conservative
component manifests itself as an exponential decay of the magnitude of the ACFs and CCFs.

When $\Omega = 0$, the movement of the probe along the $x$- and $y$-directions becomes independent. The ACFs behave as
$r_{xx} (\Delta t) = D e^{-\phi_x |\Delta t|}/\phi_x$ and
$r_{yy} (\Delta t) = D e^{-\phi_y |\Delta t|}/\phi_y$, while
the CCFs are null, $r_{xy} (\Delta t) = r_{yx} (\Delta t) = 0$. In
Fig. \ref{f:stability} this case is represented by the line
$\Omega = 0$.

When both $\Omega$ and $\Delta\phi$ are zero, the ACFs are
$r_{xx} (\Delta t) = r_{yy} (\Delta t)= D e^{-\phi |\Delta
t|}/\phi$,  and the CCFs are null, $r_{xy} (\Delta t) = r_{yx}
(\Delta t) = 0$. The corresponding force field vectors point
towards the center and are rotationally symmetric.

It is also interesting to consider the intermediate cases. In
these cases the effective angular frequency that enter the
expression is given by $\sqrt{|\Delta\phi|^2- \Omega^2}$. This
shows that the difference in the stiffness coefficients along the
$x$- and $y$-axes effectively influences the rotational term, if
this is present. A limiting case is when $|\Omega|=\Delta\phi$.
This case presents a kind of resonance between the rotational term
and the stiffness difference. However, it is not a dramatic
resonance, as it is shown by the corresponding force field (Fig. \ref{f:force-fields}).

\subsubsection{Correlation matrix in a generic coordinate system}

\begin{figure}[tbp]
\includegraphics[width = 8.5cm]{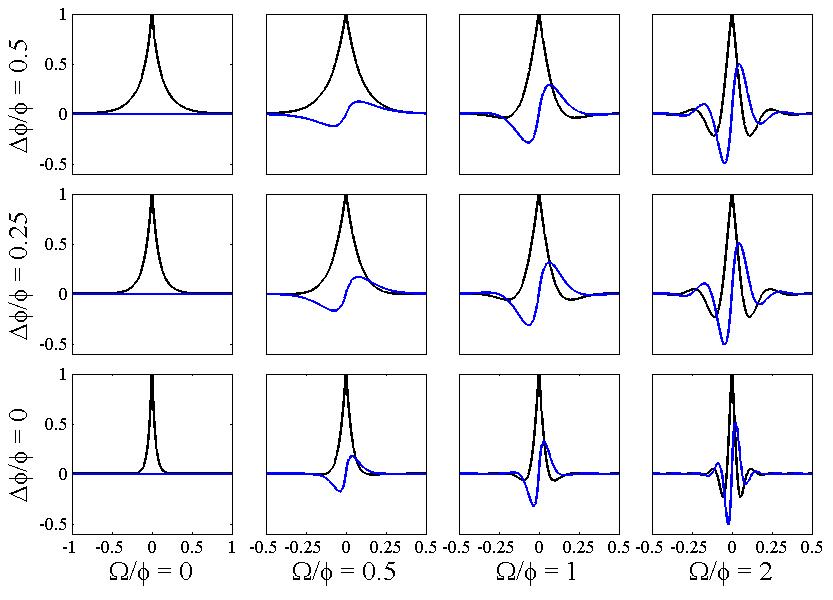}
\caption{(Color online) Functions independent from choice of the reference system for various values of the parameters
$\Delta\phi/\phi$ and $\Omega/\phi$: $r_{x'x'}+r_{y'y'}$ (black line) and
$r_{x'y'}-r_{y'x'}$ (blue line).}
\label{f:invariant}
\end{figure}

The expression for the ACFs and CCFs (\ref{e:rxx}) to (\ref{e:ryx}) were obtained in a specific coordinate system,
where the conservative and rotational component of the force field can be easily identified.
However, typically the experimentally acquired time-series of the probe position required for the calculation of the ACFs and CCFs
are given in a different coordinate system, rotated with respect to the one considered above.
If a rotated coordinate system is introduced, such that:
\begin{equation}\label{e:rotation}
\left[ {\begin{array}{*{20}c}
   {x'}  \\
   {y'}  \\
\end{array}} \right] = \left[ {\begin{array}{*{20}c}
   {\cos \theta } & { - \sin \theta }  \\
   {\sin \theta } & {\cos \theta }  \\
\end{array}} \right]\left[ {\begin{array}{*{20}c}
   x  \\
   y  \\
\end{array}} \right] ,
\end{equation}
the correlation functions in the new system are obtained as linear
combinations of (\ref{e:rxx})-(\ref{e:ryx}):
\begin{eqnarray}
r_{x'x'} (\Delta t) &=& \left(\cos{\theta}\right)^2 r_{xx} (\Delta t) - \cos{\theta} \sin{\theta} r_{xy} (\Delta t) - \sin{\theta} \cos{\theta} r_{yx} (\Delta t) + \left(\sin{\theta}\right)^2 r_{yy} (\Delta t) ,
\label{e:rxxp}
\\
r_{y'y'} (\Delta t) &=& \left(\sin{\theta}\right)^2 r_{xx} (\Delta t) + \sin{\theta} \cos{\theta} r_{xy} (\Delta t) + \cos{\theta} \sin{\theta} r_{yx} (\Delta t) + \left(\cos{\theta}\right)^2 r_{yy} (\Delta t) ,
\label{e:ryyp}
\\
r_{x'y'} (\Delta t) &=& \cos{\theta} \sin{\theta} r_{xx} (\Delta t) + \left(\cos{\theta}\right)^2 r_{xy} (\Delta t) - \left(\sin{\theta}\right)^2 r_{yx} (\Delta t) - \sin{\theta} \cos{\theta} r_{yy} (\Delta t) ,
\label{e:rxyp}
\\
r_{y'x'} (\Delta t) &=& \sin{\theta} \cos{\theta} r_{xx} (\Delta t) - \left(\sin{\theta}\right)^2 r_{xy} (\Delta t) + \left(\cos{\theta}\right)^2 r_{yx} (\Delta t) - \cos{\theta} \sin{\theta} r_{yy} (\Delta t) ,
\label{e:ryxp}
\end{eqnarray}
which in general depend on $\theta$. However, it is remarkable that the difference of the two CCFs and the
sum of the ACFs are invariant:
\begin{eqnarray}
\label{e:rxyp-ryxp}
r_{x'y'} (\Delta t) - r_{y'x'} (\Delta t)
& = &
2D \frac{e^{-\phi \left|\Delta t\right|}}{\phi}
\frac{\Omega}{\phi} \mathcal{S} (\Delta t)
, \\
\label{e:rxxp+ryyp}
r_{x'x'} (\Delta t) + r_{y'y'} (\Delta t)
& = &
2D \frac{e^{-\phi \left|\Delta t\right|}}{\phi}
\left[
\left(
\frac{ \Omega^2 - \alpha^2 \Delta\phi^2 }
{\Omega^2-\Delta\phi^2}
\right) \mathcal{C}(\Delta t)
+
\alpha^2
\frac{\Delta\phi^2}{\phi^2} \mathcal{S} (|\Delta t|)
\right]
.
\end{eqnarray}
These functions are presented in Fig. \ref{f:invariant}. These functions
are very similar to the ones presented in Fig. \ref{f:correlations};
however, the latter depend on the coordinate system choice.

\begin{figure}[tbp]
\includegraphics[width = 8.5cm]{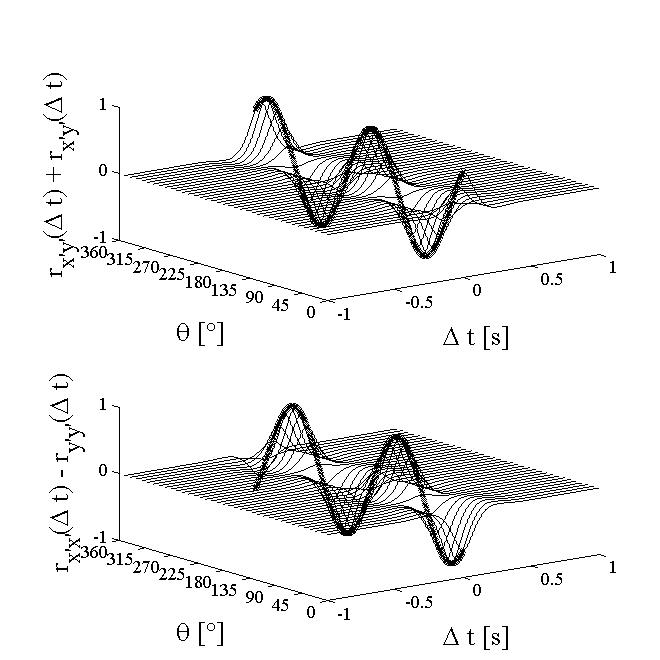}
\caption{ (Color online) Functions which depend on the orientation of the coordinate system for various values of the parameters
as a function of the angle with respect to the coordinate system chosen in the section \ref{sec:theory}: (a) $r_{x'x'}-r_{y'y'}$ and (b)
$r_{x'y'}+r_{y'x'}$. The thicker lines show the values for $\Delta t = 0$.}
\label{f:variant}
\end{figure}

Other two combinations of (\ref{e:rxxp})-(\ref{e:ryxp}), which are also useful for the analysis of the experimental
data, namely the sum of the CCFs and the difference of the ACFs,
depend on the choice of the
reference frame:
\begin{eqnarray}
\label{e:rxyp+ryxp}
r_{x'y'} (\Delta t) + r_{y'x'} (\Delta t)
& = &
2D \frac{e^{-\phi \left|\Delta t\right|}}{\phi} \alpha^2
\frac{\Delta \phi}{\phi}
\left( \mathcal{C} (\Delta t) + \mathcal{S} (|\Delta t|) \right)
\left( \frac{\Omega}{\phi} \cos{(2\theta)} - \sin{(2\theta)} \right)
, \\
\label{e:rxxp-ryyp}
r_{x'x'} (\Delta t) - r_{y'y'} (\Delta t) & = & -2D
\frac{e^{-\phi \left|\Delta t\right|}}{\phi} \alpha^2
\frac{\Delta \phi}{\phi} \left( \mathcal{C} (\Delta t) +
\mathcal{S} (|\Delta t|) \right) \left( \frac{\Omega}{\phi}
\sin{(2\theta)} + \cos{(2\theta)} \right) .
\end{eqnarray}
Their plots are shown in Fig. \ref{f:variant}. In particular, when they are
evaluated for $\Delta t = 0$, they deliver information on the orientation of
the coordinate system.

\subsubsection{Power Spectral Density Matrix}

In the frequency domain the equation (\ref{e:ForceFieldBM}) is given by:
\begin{equation}
\label{} i 2 \pi f\mathbf{R}(f) = \mathbf{J_0} \mathbf{R}(f) +
\sqrt{2 D} \mathbf{H}(f),
\end{equation}
and its solution is $\mathbf{R}(f) = \sqrt{2D} \left(i 2 \pi f
\mathbf{I_2} - \mathbf{J_0}\right)^{-1} \mathbf{H}(f)$, where
$\mathbf{I_2}$ is the 2D unit matrix, and the
corresponding PSD matrix:
\begin{equation}
\label{} \mathbf{P}(f) = \mathbf{R}\cdot\mathbf{R^h} =
\frac{2D}{|\left(\phi_x + i 2 \pi f\right)\left(\phi_y + i 2
\pi f\right)+\psi^2|^2} \left[\begin{array}{cc} \phi_y^2 + 4
\pi^2 f^2 + \psi^2 & \psi \left[ \phi_x - \phi_y - i 4 \pi f
\right]
\\
\psi \left[ \phi_x - \phi_y + i 4 \pi f \right] &
\phi_x^2 + 4 \pi^2 f^2 + \psi^2
\\
\end{array}\right]
\end{equation}
where the property
$\mathbf{H}(f) \cdot \mathbf{H^h}(f) = \mathbf{I_2}$ has been
used. We notice that the PSD matrix could have been obtained as Fourier-transfor of the
correlation matrix (Wiener-Khintchine theorem).

\section{Experimental Considerations}\label{sec:experiments}

In this section we propose a concrete analysis workflow to
reconstruct the force field from the experimental time-series of
the probe position.

Experimentally
the probe position time-series is the only available
information to reconstruct the force field. Typically these data
are obtained in an arbitrary coordinate system. These time-series need to
be statistically analyzed in order to reconstruct all the
parameters of the force field, i.e. $\phi$, $\Delta\phi$,
and $\Omega$, and the orientation of the coordinate system. The
detailed procedure to retrieve all this information from the
experimental data is presented in this section.

Let us suppose to have the probe position time-series in a generic
coordinate system $\mathbf{r'}(t) = \left[ x'(t), y'(t) \right]$,
First, we evaluate the parameters $\phi$, $\Delta\phi$, and $\Omega$.
Then, we transform the coordinate system to
the one presented in the section \ref{sec:theory}, where the conservative and
rotational components are separated.
Finally, we reconstruct the total force field.
Eventually, the trapping force field may be subtracted
to retrieve the external force field under investigation.

\begin{figure}[tbp]
\includegraphics[width = 8.5cm]{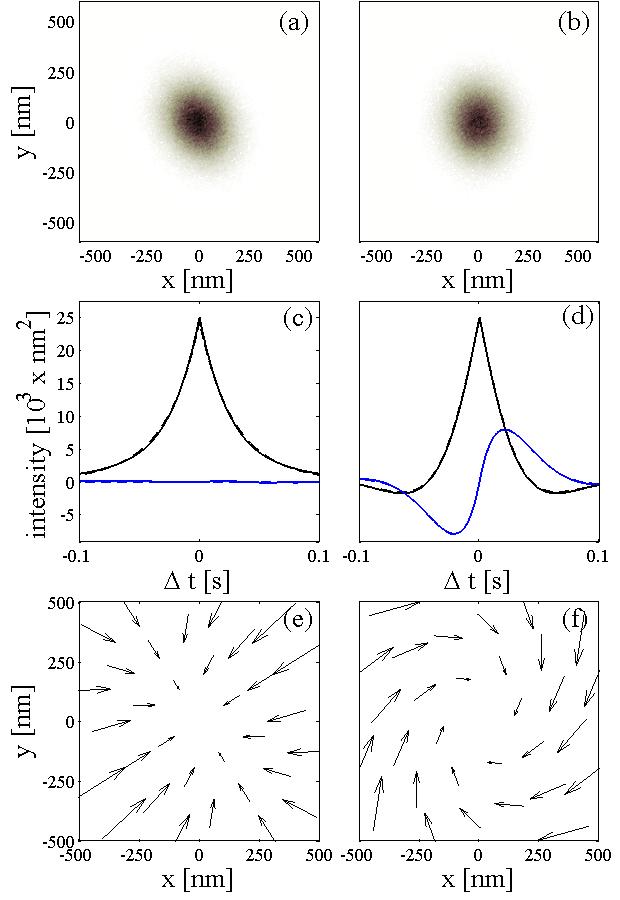}
\caption{ (Color online) (a-b) Probability density function for a Brownian particle under the influence of the force-field shown in the inset; in (a) the
force-field is purely conservative, while in (b) it has a rotational component. (c-d) Invariant function,
$r_{x'x'}+r_{y'y'}$ (black line) and $r_{x'y'}-r_{y'x'}$ (blue line). (e-f) Force fields obtained form the numerically simulated data.}
\label{f:simulations}
\end{figure}

In order to illustrate this method we proceed to analyze some
numerically simulated data. The main steps of this analysis are
presented in Fig. \ref{f:simulations}.
In Fig. \ref{f:simulations}(a) the PDF is
shown for the case of a probe in a force field with the
following parameters: $\phi = 37.1\, s^{-1}$, $\Delta\phi =
9.3\, s^{-1}$ (corresponding to $k_x = 43.25\,pN/\mu m$ and $k_y = 26.25\,pN/\mu m$), $\Omega = 0$, and $\theta = 30^{\circ}$. The
PDF is ellipsoidal due to the difference of the stiffness along
two orthogonal directions. In Fig.
\ref{f:simulations}(b) the PDF for a force field with the same $\phi$ and $\Delta\phi$
but with $\Omega = 37.1 \, s^{-1}$ is presented. The presence of the
rotational component in the force field produces two main effects.
First, the PDF is more rotationally-symmetric and its main axes
undergo a further rotation. Secondly, as we show below, the CCF is not null (Fig.
\ref{f:simulations}(d)).

\subsection{Evaluation of the parameters $\phi$, $\Delta\phi$, and $\Omega$}

In order to evaluate the parameters of the force field, $\phi$,
$\Delta\phi$, and $\Omega$, we calculate the
correlation matrix in the coordinate system where the experiments have been done,
\begin{equation}\label{sim1}
\left\langle {{\bf r'}(\Delta t){\bf r'}^{\bf h} (0)}
\right\rangle  = \left( {\begin{array}{*{20}c}
r_{x'x'} (\Delta t) & r_{x'y'} (\Delta t) \\
r_{y'x'} (\Delta t) & r_{y'y'} (\Delta t) \\
\end{array}} \right).
\end{equation}
Then we calculate the CCF difference (\ref{e:rxyp-ryxp}), $r_{x'y'} (\Delta t) - r_{y'x'} (\Delta t)$.
As we showed in section II, this function is
invariant with respect to the choice of the reference system, and
it is different form zero only if $\Omega \neq 0$. The results are shown in Fig.
\ref{f:simulations}(c) and \ref{f:simulations}(d) for the
cases of the data shown in Fig. \ref{f:simulations}(a) and \ref{f:simulations}(b)
respectively. The three aforementioned parameters can
be found by fitting the experimental data to the theoretical shape
of this function. In particular, the exponential decay of the
function is related to the $\phi$ parameter; the period of the
superimposed oscillations is related to the effective angular
frequency $\sqrt{|\Delta\phi^2-\Omega^2|}$; and the sign of the
slope in $\Delta t = 0$ gives the sign of $\Omega$.

When $\Omega = 0$, the CCF difference (\ref{e:rxyp-ryxp})
is null (Fig. \ref{f:simulations}(c)), it can not be used to find the
two remaining parameters. For $\Omega = 0$, the other invariant function,
the ACF sum (equation (\ref{e:rxxp+ryyp})), is given by
\begin{equation}\label{sim2}
r_{x'x'} (\Delta t) + r_{y'y'} (\Delta t)
=
2D \frac{e^{-\phi \left|\Delta t\right|}}{\phi}
\left[
\frac{\phi^2}{\phi^2 - \Delta\phi^2}
\mathcal{C}(\Delta t)
+
\frac{\Delta\phi^2}{\phi^2 - \Delta\phi^2}
\mathcal{S}(\Delta t)
\right] .
\end{equation}
$\phi$ and $\Delta\phi$ can be evaluated
by fitting the data to (\ref{sim2}). The function (\ref{e:rxxp+ryyp})
can be used for the fitting of the three parameters but can
not give information on the sign of $\Omega$, which must be
retrieved from the sign of the slope at $\Delta t = 0$ of the CCF difference.

\subsection{Coordinate system transformation}

Although the values of the parameters $\phi$, $\Delta\phi$, and $\Omega$ are now known,
the directions of the force vectors are still
missing.
In order to retrieve the orientation of the experimental coordinate system,
we now use the orientation dependent functions (\ref{e:rxyp+ryxp}) and (\ref{e:rxxp-ryyp}).
The best choice is
to evaluate the two functions for $\Delta t = 0$,
because the signal-to-noise ratio is highest at this point:
\begin{equation}\label{sim4}
    \left\{ {\begin{array}{*{20}c}
   {r_{x'y'} (0) + r_{y'x'} (0)= 2D \frac{\alpha^2}{\phi} \frac{\Delta \phi}{\phi}
   \left( \frac{\Omega}{\phi} \cos{(2\theta)} - \sin{(2\theta)}\right) }
    \\
   {r_{x'x'} (0) - r_{y'y'} (0)= -2D \frac{\alpha^2}{\phi} \frac{\Delta \phi}{\phi}
   \left( \frac{\Omega}{\phi} \sin{(2\theta)} + \cos{(2\theta)}\right)}
    \\
    \end{array}} \right . .
\end{equation}
The solution of this system delivers the value of the rotation angle $\theta$.
If $\Delta \phi = 0$, (\ref{sim4}) is undetermined as a
consequence of the PDF radial symmetry. In this case any orientation can be used.
If $\Omega = 0$, the orientation of the coordinate system coincide with the
axis of the PDF ellipsoid and, although (\ref{sim4}) can still be used,
the Principal Component Analysis (PCA) algorithm is a convenient means to determine their directions.

\subsection{Reconstruction of the force field}

Now everything is ready to reconstruct the unknown force field
acting on the probe around the equilibrium
position in an area comparable with the mean square displacement of the
probe. From the values of $\phi$ and $\Delta\phi$, the conservative forces acting on the probe
result ${\bf f}_c(x,y) =  - \left( {k_x x{\bf e}_x  + k_y y{\bf e}_y}\right)$ and,
from the values of $\Omega$, the rotational force is
${\bf f}_r(x,y) =  \Omega\left( {y{\bf e}_x  - x{\bf e}_y}\right)$.
The total force field is
\begin{equation}\label{}
    {\bf f}(x,y) =  {\bf f}_c(x,y) +{\bf f}_r(x,y) =
    \left(-k_x x + \Omega y\right){\bf e}_x  - \left(k_y y + \Omega x\right){\bf e}_yx
\end{equation}
in the rotated coordinate system. Now the rotation (\ref{e:rotation}) can be used
in order to have the force field in the experimental coordinate system.
The unknown component can be easily
reconstructed by subtraction of the know ones, such as the optical
field.

\section{Experimental results}\label{sec:cases}

\begin{figure}[tbp]
\includegraphics[width = 8.5cm]{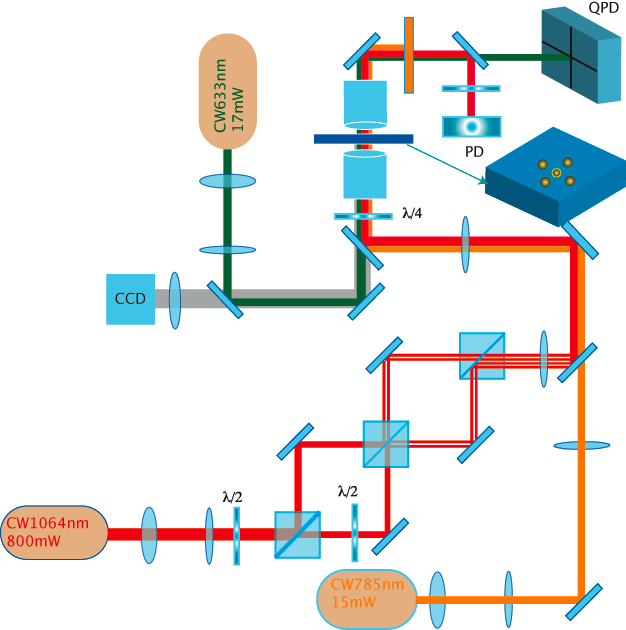}
\caption{ (Color online) Experimental setup.}
\label{f:setup}
\end{figure}

For an experimental verification of our conclusions, we
analyze the Brownian motion of an optically trapped
polystyrene sphere in the presence of an external force field generated by a fluid flow \cite{Giorgio2007}.
A schematic of the setup is presented in Fig. \ref{f:setup}.

\begin{figure}[tbp]
\includegraphics[width = 8.5cm]{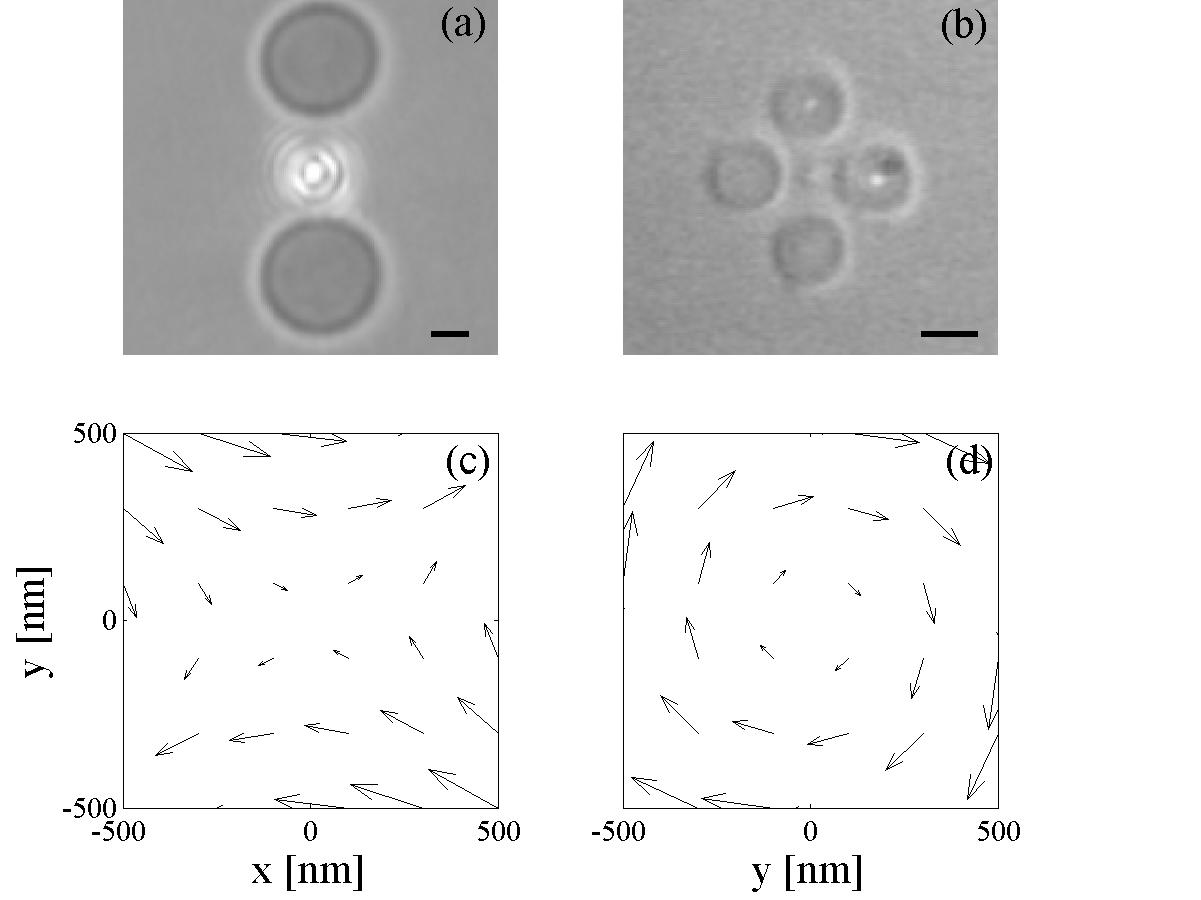}
\caption{ (Color online) Experimental configuration with
two (a) and four (b) spinning beads and respective force fields (c) and (d).}
\label{f:setup2}
\end{figure}

An optical trap is generated
by a CW $633\, nm$ beam at the focal plane of a $100 \times$  $1.3 NA$
objective lens inside a chamber. The chamber is
prepared using two cover slips separated by a $50\, \mu m$
spacer and filled with a solution containing polystyrene
spheres (radius $R=0.5\,\mu m$). The forward scattered light
from the trapped sphere is collimated by a $50\times$  objective
onto a quadrant photodiode (QPD). The trap force constant
can be adjusted by changing the intensity of the laser
beam.

The fluid flow that produces the external force field was generated
using solid spheres made of a birefringent material (Calcium
Vaterite Crystals (CVC) spheres, radius $R = 1.5 \pm 0.2 \mu m$
\cite{Bishop2004}), which can be made spin due to the transfer of
orbital angular momentum of light. They are all-optically
controlled, i.e. their position can be controlled by an optical
trap and their spinning state can be controlled through the
polarization state of the light. In our
experimental realization up to four CVC spheres were optically
trapped in water and put into rotation using four steerable
$1064\,nm$ beams from a Nd:YAG laser with controllable
polarization - to control the direction of the rotation - and
power - to control the rotation rate.

\subsection{Conservative force field}

In order to produce a conservative force field, two CVC were placed as shown in
Fig. \ref{f:setup2}(a), which should theoretically produce the force field presented in
Fig. \ref{f:setup2}(c).

\begin{figure}[tbp]
\includegraphics[width = 8.5cm]{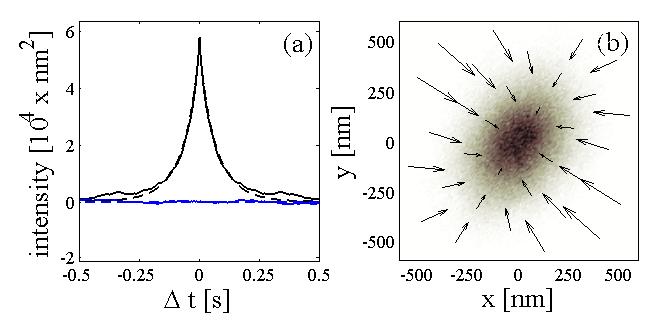}
\caption{ (Color online)
(a) Invariant functions, $r_{x'x'}+r_{y'y'}$ (black line) and $r_{x'y'}-r_{y'x'}$ (blue line),
and respective fitting to the theoretical shape (dotted lines).
(b) Experimental probability density function and estimated force field.}
\label{f:two}
\end{figure}

In Fig. \ref{f:two}(a), the invariant functions,
$r_{x'x'}+r_{y'y'}$ (black line) and $r_{x'y'}-r_{y'x'}$ (blue line),
and respective fitting to the theoretical shapes are presented.
The CCF difference tells us that $\Omega = 0$ in this case,
while the fitting to the ACF sum tells us the values of $\phi = 18 \, s^{-1}$ and $\Delta\phi = 6 \, s^{-1}$.
The value of the rotation of the coordinate system in this case is $32^{\circ}$.

The total force field can now be recontructed: $k_x = 225\, fN/\mu m$ and $k_y = 112 \, fN/\mu m$.
This force field is presented in Fig. \ref{f:two}(b).
We can now retrieve the
hydrodynamic force field by subtracting the optical force field ($k_{opt} = 185 \, fN/ \mu m$
approximatively constant in all directions), that
can be measured in absence of rotation of the spinning beads.

\subsection{Rotational force field}

In order to produce a rotational force field, four CVC were placed as shown in
Fig. \ref{f:setup2}(b), which should theoretically produce the force field presented in
Fig. \ref{f:setup2}(d).

\begin{figure}[tbp]
\includegraphics[width = 8.5cm]{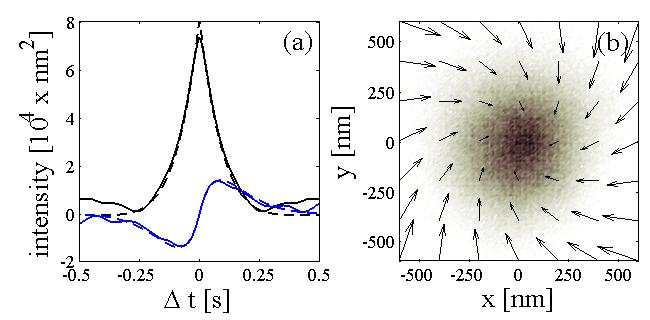}
\caption{ (Color online)
(a) Invariant functions, $r_{x'x'}+r_{y'y'}$ (black line) and $r_{x'y'}-r_{y'x'}$ (blue line),
and respective fitting to the theoretical shape (dotted lines).
(b) ) Experimental probability density function and estimated fore field.}
\label{f:four}
\end{figure}

In Fig. \ref{f:four}(a), the invariant functions,
$r_{x'x'}+r_{y'y'}$ (black line) and $r_{x'y'}-r_{y'x'}$ (blue
line), and respective fitting to the theoretical shapes are
presented. Now the CCF difference is not null any more and
tehrefore it can be used to fit the three parameters: $\phi = 11\,
s^{-1}$, $\Delta \phi \approx 0$, and $\Omega = 5\, rads^{-1}$. We
can notice that the ACF sum can be used for this purpose as well;
however, we have to remark that using the latter the sign of
$\Omega$ stays undetermined. The small value of $\Delta\phi$
implicates that the rotation of the coordinate system is not
crucial.

The total force field can now be recontructed: $k_x \approx k_y = 100 \, fN/\mu m$.
This force field is presented in Fig. \ref{f:four}(b).
We can now retrieve the
hydrodynamic force field by subtracting the optical force field ($k_{opt} = 78 \, fN/\mu m$
approximatively constant in all directions), that
can be measured in absence of rotation of the spinning beads.

\section{Conclusion}

We have shown how the PFM can be applied to the detection of locally non-homogeneous force fields.
This has been achieved by analyzing the ACFs and CCFs of the probe position time-series.
We believe that this technique can gain new insights into micro- and molecular-scale phenomena. In these cases
the presence of the Brownian motion is intrinsic and has can not be disregarded. Therefore this technique permits one to
take advantage to the Brownian fluctuations of the probe in order to explore the force field present in its surroundings.

One of the most remarkable advantages of the technique we propose is that it can be implemented in all
existing PFM-setups and even on data acquired in the past. Indeed, it does not require changes to be made in the physical setup,
but only to analyze the data in a deeper way.

\begin{acknowledgments}
The authors acknowledge useful discussions with N. Heckenberg, A.
Bagno, and M. Rub\'{i}.
This research was carried out in the framework of ESF/PESC (Eurocores
on Sons), through grant 02-PE-SONS-063-NOMSAN, and with the
financial support of the Spanish Ministry of Education and
Science. It was also partially supported by the Departament
d'Universitats, Recerca i Societat de la Informaci\'{o} and the
European Social Fund.
\end{acknowledgments}


\end{document}